\documentclass[10pt,american,preprint, aps, nofootinbib]{revtex4-1}
\usepackage[T1]{fontenc}
\usepackage[latin9]{inputenc}
\usepackage{babel}
\usepackage{mathrsfs}
\usepackage{amsmath}
\usepackage{amssymb}
\usepackage{graphicx}
\usepackage{wasysym}
\usepackage[pdfusetitle,
 bookmarks=true,bookmarksnumbered=false,bookmarksopen=false,
 breaklinks=false,pdfborder={0 0 1},backref=false,colorlinks=false]
 {hyperref}

\makeatletter

\usepackage{etoolbox}
\usepackage{breqn}

\makeatletter
\let\cat@comma@active\@empty
\makeatother

\newcommand{\breqnoverloadothers}
{%
    \renewenvironment{equation}{\ignorespaces\begin{dmath}}{\end{dmath}\ignorespacesafterend}%
    \renewenvironment{equation*}{\ignorespaces\begin{dmath*}}{\end{dmath*}\ignorespacesafterend}%
    \renewenvironment{multline}{\ignorespaces\begin{dmath}}{\end{dmath}\ignorespacesafterend}%
    \renewenvironment{multline*}{\ignorespaces\begin{dmath*}}{\end{dmath*}\ignorespacesafterend}%

}

\newcommand\breqnundefineothers
{%
    \renewenvironment{equation}{}{}%
    \renewenvironment{equation*}{}{}%
    \renewenvironment{multline*}{}{}%

}

\AtBeginEnvironment{dmath}{\breqnundefineothers}
\AtBeginEnvironment{dmath*}{\breqnundefineothers}

\AtBeginEnvironment{dgroup}{\def\breqnundefineothers{}\breqnoverloadothers}
\AtBeginEnvironment{dgroup*}{\def\breqnundefineothers{}\breqnoverloadothers}

\newcommand\brwrap[3]{%
  \setbox0=\hbox{$#2$}
  \left#1\vbox to \the\ht0{\hbox to 0pt{}}\right.\kern-.2em
  \begingroup #2\endgroup\kern-.15em
  \left.\vbox to \the\ht0{\hbox to 0pt{}}\right#3
}

\makeatother

\begin{document}
\title{Dynamical Black Hole Emission}
\author{David A. Lowe}
\affiliation{Department of Physics, Brown University, Providence, RI 02912, USA}
\author{Larus Thorlacius}
\affiliation{Science Institute, University of Iceland, Dunhaga 3, 107 Reykjavik,
Iceland}
\begin{abstract}
Semiclassical black hole emission in four spacetime dimensions is
studied using a non-local effective action. The field equations that
determine the time-dependent renormalized stress tensor are solved
numerically for a black hole formed by an ingoing null shock wave,
and otherwise smooth initial data. We find that Hawking radiation
is generated dynamically near the black hole horizon and freely propagates
out to null infinity, resulting in an outgoing energy flux that builds
up from zero at early retarded times before the black hole forms.
This resolves the long-standing issue of pre-Hawking radiation, suffered
by calculations based on a static approximation to the stress tensor
in an Unruh state, and paves the way towards four-dimensional black
hole evolution with semiclassical back-reaction included.
\end{abstract}
\maketitle

\section{Introduction}

Hawking\textquoteright s prediction \citep{Hawking1975} that gravitational
collapse produces late-time thermal radiation remains a cornerstone
of semiclassical gravity. The derivation, based on quantum field theory
in curved spacetime, reveals that black holes emit a steady flux at
future null infinity, yet leaves open difficult questions of how the
radiation turns on dynamically and how back-reaction modifies the
spacetime over time. Substantial analytic progress in understanding
semiclassical evaporation has been achieved in two spacetime dimensions,
where the CGHS and RST models \citep{Callan:1992rs,Russo:1992ax}
provide solvable settings in which the trace anomaly can be rewritten
in local form and the full time-dependent stress tensor can be computed.
The two-dimensional results show that Hawking flux vanishes prior
to horizon formation, then rises gradually to a steady value once
a trapped region forms, thereby demonstrating that radiative output
is generated dynamically rather than imposed at past infinity.

In four spacetime dimensions, the anomaly-induced part of the effective
action is more intricate. The covariant non-local anomaly functional
introduced by Riegert \citep{RIEGERT198456} can nevertheless be rewritten
equivalently as a local fourth-order scalar--tensor theory by the
introduction of an auxiliary scalar field. This anomaly-induced effective
field theory has been studied in the static Schwarzschild geometry
to obtain the semiclassical stress tensor \citep{Balbinot:1999ri,Mottola:2006ew,Mottola:2016aa,Mottola:2025fhl,Lowe:2025bxl}.
In particular, in recent work, we have shown how regularity at the
future horizon and vanishing ingoing energy flux at past null infinity
select a unique Unruh-like configuration with no scalar hair and finite
late-time luminosity \citep{Lowe:2025bxl}. However, because that
construction assumes a static stress tensor, it inevitably produces
an unphysical \textquotedblleft pre-Hawking\textquotedblright{} flux
before the onset of collapse, a feature that must be resolved in any
truly dynamical calculation. Thus, a dynamical implementation of the
field equations in a collapsing geometry is required to eliminate
pre-formation radiation and to determine the onset of Hawking emission
from evolution. 

The primary goal of the present paper is to demonstrate for the first
time that an effective field theory approach in 3+1 spacetime dimensions,
where Einstein gravity is supplemented by higher order corrections
that reproduce the matter induced trace anomaly, successfully captures
the dynamical onset of Hawking emission. We obtain the following main
results:
\begin{itemize}
\item Uniqueness from smooth initial data: Starting from regular data on
$\mathscr{I}^{-}$, the dynamics fix the flux uniquely without requiring
additional boundary prescriptions---consistent with the static uniqueness
result of Unruh state \citep{Lowe:2025bxl}, but now achieved dynamically.
\item Gradual turn-on of Hawking radiation: The outgoing flux measured at
$\mathscr{I}^{+}$ vanishes at early retarded time and rises smoothly
to a constant plateau only after horizon formation, matching expectations
from two-dimensional semiclassical models, but now demonstrated in
four dimensions.
\item Numerical control of the higher-derivative field equations: The solution
demonstrates that dynamical semiclassical evolution in 3+1 dimensions
is tractable, opening the way for future incorporation of back-reaction
and evaporation-driven metric evolution.
\end{itemize}

\section{Exact results in a two-dimensional model}

As a warmup exercise, let us analyze the onset of Hawking emission
in the simplified context of the CGHS model of two-dimensional dilaton
gravity \citep{Callan:1992rs}. The motivation is twofold. First of
all, it serves to illustrate the auxiliary field method that we employ
for the more challenging four-dimensional case in a later section
of the paper. Second, it provides a closed form expression for the
outgoing energy flux of Hawking radiation as a function of retarded
time at future null infinity. As we see below, our numerical results
for Hawking emission in four spacetime dimensions follow a qualitatively
similar pattern. 

The CGHS model has classical black hole solutions with the same global
causal structure as asymptotically flat black holes in 3+1-dimensional
Einstein gravity. In particular, there are dynamical solutions describing
black holes formed by the gravitational collapse of matter into a
smooth initial vacuum configuration. The classical action is 
\begin{equation}
S=\frac{1}{2\pi}\int d^{2}x\sqrt{-g}e^{-2\varphi}\left(R+4(\nabla\varphi)^{2}+4\lambda^{2}\right)+S_{CFT}\,,\label{eq:cghsaction}
\end{equation}
where $\lambda$ is a characteristic length that can be set to $\lambda=1$
by rescaling the two-dimensional coordinates. $S_{CFT}$ is the action
of conformally coupled matter fields with large central charge $c\gg24$
to justify a semiclassical treatment. 

We adapt a conformal gauge for the two-dimensional metric,
\begin{equation}
ds^{2}=-e^{-2\rho(x^{+},x^{-})}dx^{+}dx^{-},\label{eq:confgauge}
\end{equation}
and choose so-called Kruskal coordinates, where the conformal factor
is equal to the dilaton field, $\rho=\varphi$. Then the classical
field equations take a particularly simple form,
\begin{equation}
\partial_{+}\partial_{-}e^{-2\varphi}=-1\,,\qquad-2\partial_{\pm}^{2}e^{-2\varphi}=T_{\pm\pm},\label{eq:fieldeqs}
\end{equation}
where $T_{\pm\pm}$ are the non-trivial components of the matter stress-energy
tensor in conformal gauge. The following solution describes a black
hole formed by an infinitely thin ingoing null shockwave at $x^{+}=x_{0}^{+}$,
\begin{equation}
e^{-2\rho}=e^{-2\varphi}=\begin{cases}
-x^{+}x^{-} & \text{if}\qquad x^{+}<x_{0}^{+}\,,\\
M-x^{+}(x^{-}+\frac{M}{x_{0}^{+}}) & \text{if}\qquad x^{+}>x_{0}^{+}\,.
\end{cases}\label{eq:bhsol}
\end{equation}
In the $x^{+}<x_{0}^{+}$ region, the following change of coordinates
\begin{equation}
x^{+}=e^{\omega^{+}},\qquad x^{-}=-e^{-\omega^{-}},\label{eq:vacflat}
\end{equation}
brings the metric to a manifestly Minkowski form, with $\rho(\omega^{+},\omega^{-})=0$.
Meanwhile, 
\begin{equation}
x^{+}=e^{\sigma^{+}},\qquad x^{-}=-\frac{M}{x_{0}^{+}}-e^{-\sigma^{-}},\label{eq:asymptoticflat}
\end{equation}
defines asymptotically Minkowski coordinates in the $x^{+}>x_{0}^{+}$
region outside the shockwave.

A two-dimensional analog of the Hawking effect arises when quantum
effects of matter fields are considered in this classical black hole
background. The standard calculation \citep{Callan:1992rs,Christensen:1977jc},
obtains the outgoing energy flux in closed form by starting with the
conformal anomaly, 
\begin{equation}
\left\langle T_{a}^{\:a}\right\rangle =\frac{c}{12}R\,,\label{eq:2d_anomaly}
\end{equation}
and integrating the conservation equations $\nabla^{a}T_{ab}=0$ in
conformal gauge, while imposing suitable boundary conditions at past
null infinity and in the vacuum inside the ingoing null shell. This
method is not applicable in four spacetime dimensions, where the general
form for the conformal anomaly is more involved than \eqref{eq:2d_anomaly}
and there is no analog of the conformal gauge. We can instead adopt
an effective field theory approach that yields the same closed form
expression for the outgoing Hawking flux but also generalizes to higher
dimensions in a straightforward manner.

The one-loop contribution to the quantum effective action due to the
matter fields is given by the non-local Polyakov-Liouville term \citep{POLYAKOV1981207},
\begin{equation}
S_{PL}=-\frac{c}{96\pi}\int d^{2}x\sqrt{-g(x)}\int d^{2}y\sqrt{-g(y)}R(x)G(x,y)R(y)\,,\label{eq:Polyakovterm}
\end{equation}
where $G(x,y)$ is a Green function for the scalar Laplacian, $\sqrt{-g(x)}\nabla_{x}^{2}G(x,y)=\delta^{(2)}(x,y)$.
The original formulation of the semiclassical CGHS model \citep{Callan:1992rs}
took advantage of the fact that this term appears local in conformal
gauge. More generally, the effective action can be rendered local
by introducing an auxiliary scalar field that is sourced by the scalar
curvature, 
\begin{equation}
S_{Z}=\frac{c}{48\pi}\int d^{2}x\sqrt{-g}\left(-\frac{1}{2}(\nabla Z)^{2}+R\,Z\right).\label{eq:Zaction}
\end{equation}
The non-local form \eqref{eq:Polyakovterm} is recovered by eliminating
the auxiliary scalar by its field equation, 
\begin{equation}
\nabla^{2}Z=-R\,.\label{eq:Zeq}
\end{equation}
Quantum effects of the original matter fields are encoded in the induced
stress-energy tensor,
\begin{equation}
T_{ab}=-\frac{4\pi}{\sqrt{-g}}\frac{\delta S_{Z}}{\delta g^{ab}}=\frac{c}{24}\left(\nabla_{a}Z\,\nabla_{b}Z+2\nabla_{a}\nabla_{b}Z-g_{ab}\left(\frac{1}{2}(\nabla Z)^{2}+2\nabla^{2}Z\right)\right).\label{eq:2d_stress}
\end{equation}
evaluated on shell for a solution of the scalar field equation \eqref{eq:Zeq}
in the black hole background \eqref{eq:bhsol}. In particular, the
outgoing Hawking energy flux is given by the $T_{--}$ component of
the stress-energy tensor, evaluated at future null infinity in the
asymptotically Minkowski coordinates \eqref{eq:asymptoticflat}. In
these coordinates, the scalar field equation reduces to 
\begin{equation}
\partial_{+}\partial_{-}Z=\frac{M\,e^{\sigma^{-}-\sigma^{+}}}{(1+M\,e^{\sigma^{-}-\sigma^{+}})^{2}},\label{eq:Zeq_outside}
\end{equation}
and integrating once gives
\begin{eqnarray}
\partial_{-}Z(\sigma^{+},\sigma^{-}) & = & \int_{\sigma_{0}^{+}}^{\sigma^{+}}d\tilde{\sigma}^{+}\frac{M\,e^{\sigma^{-}-\tilde{\sigma}^{+}}}{(1+M\,e^{\sigma^{-}-\tilde{\sigma}^{+}})^{2}}\nonumber \\
 & = & \frac{1}{1+M\,e^{\sigma^{-}-\sigma^{+}}}-\frac{1}{1+M\,e^{\sigma^{-}-\sigma_{0}^{+}}}.\label{eq:dmZ}
\end{eqnarray}
We have set the lower limit of integration at $\sigma^{+}=\sigma_{0}^{+}=\log x_{0}^{+}$
to ensure that $\partial_{-}Z=0$ at $\sigma^{+}=\sigma_{0}^{+}$.
Then no energy flux emerges from the vacuum region and any outgoing
radiation at $\sigma^{+}\rightarrow\infty$ is sourced by the background
curvature outside the shockwave. Inserting into $T_{--}(\sigma^{+},\sigma^{-})=\frac{c}{24}(\nabla_{-}Z\nabla_{-}Z+2\nabla_{-}\nabla_{-}Z)$
and taking the $\sigma^{+}\rightarrow\infty$ limit gives
\begin{equation}
T_{--}\longrightarrow\frac{c}{24}\left(1-\frac{1}{(1+M\,e^{\sigma^{-}-\sigma_{0}^{+}})^{2}}\right).\label{eq:outflux}
\end{equation}
This precisely matches the original result of Callan \emph{et al.}
\citep{Callan:1992rs}. The outgoing energy flux vanishes at early
retarded times $\sigma^{-}\rightarrow-\infty$ and then makes a smooth
transition to constant outgoing flux at $\sigma^{-}\sim\sigma_{0}^{+}-\log M$.
Since the calculation does not include semiclassical back-reaction
due to Hawking emission, the outgoing flux persists to $\sigma^{-}\rightarrow+\infty$. 

\section{Basic setup in four-dimensional spacetime\label{sec:Basic-setup-in}}

We now turn our attention to the more challenging problem of Hawking
emission in four-dimensional spacetime, using a higher-derivative
effective field theory with an auxiliary scalar field playing a similar
role as the $Z$ field in the two-dimensional theory. The calculation
of the Hawking flux from a black hole formed by a spherically symmetric
ingoing null shockwave is conceptually similar to the two-dimensional
problem but the four-dimensional effective field theory is more complicated
and we have to settle for numerical solutions of the time-dependent
field equations. 

Our starting point is 3+1-dimensional Einstein gravity coupled to
Riegert's effective action \citep{RIEGERT198456}, written in local
form by introducing an auxiliary scalar field,\footnote{We follow the notation and conventions of \citep{Lowe:2025bxl} for
the remainder of the present work.} \begin{dmath}
\begin{equation}
S=\int d^{4}x\,(-g)^{1/2}\left[\frac{1}{16\pi}R+\frac{1}{192\pi^{2}}(c-\tfrac{2}{3}b)R^{2}-\tfrac{b}{2}\nabla^{2}\phi\nabla^{2}\phi-\tfrac{b}{3}R(\nabla\phi)^{2}+bR^{ab}\nabla_{a}\phi\nabla_{b}\phi+\frac{\phi}{8\pi}\left((a+b)C^{2}+\frac{2b}{3}\left(R^{2}-3R_{ab}R^{ab}-\nabla^{2}R\right)\right)\right].\label{eq:action}
\end{equation}
\end{dmath} The metric equations of motions can be rearranged to
read $R_{ab}-\frac{1}{2}R=8\pi T_{ab}$, where the semiclassical stress
tensor on the right hand side comes from the terms in \eqref{eq:action}
involving the scalar field and the $O(R^{2)}$ term. The stress tensor
was worked out in \citep{Mottola:2016aa,Lowe:2025bxl} and we will
not repeat the rather long expression here. The action \eqref{eq:action}
is constructed so that, when the scalar field is eliminated by its
equation of motion, we reproduce the trace anomaly for conformally
coupled, electrically neutral, matter in four spacetime dimensions,
\begin{equation}
g^{ab}\left\langle T_{ab}\right\rangle =\frac{1}{16\pi^{2}}\left(aC^{2}+bE-c\nabla^{2}R\right)\,,\label{eq:anomaly}
\end{equation}
with 
\begin{align}
C^{2} & =R^{abcd}R_{abcd}-2R^{ab}R_{ab}+\frac{1}{3}R^{2},\label{eq:c2e}\\
E & =R^{abcd}R_{abcd}-4R^{ab}R_{ab}+R^{2},\nonumber 
\end{align}
the square of the Weyl tensor and the Euler density, respectively.
The coefficients $a,b,c$ depend on the matter fields content of the
theory and we are interested in a semiclassical limit where the anomaly
coefficients are large, due to a large number of matter fields $N$,
but we scale $N\to\infty$ with $\hbar N$ fixed to suppress graviton
loops. The $b$ coefficient is negative for ordinary spin $\leq1$
matter fields \citep{Duff:1977ay,Duff:1993wm}. In earlier work on
black hole emission in this model \citep{Lowe:2025bxl}, we found
that this negative sign leads to a negative outgoing energy flux at
future null infinity. This can be remedied by adding to the effective
action a scale invariant term, involving the second auxiliary scalar
that couples to the square of the Weyl tensor, which contributes a
positive outgoing energy flux without affecting the trace anomaly
\citep{Liu:2025xfu}. The resulting effective field action is rather
unwieldy for numerical computations and we instead work with the simpler
action \eqref{eq:action} and impose the condition $b>0$ by hand
to ensure a positive outgoing energy flux. 

As a first step towards analyzing the time-dependent equations of
motion, we consider a black hole formed via collapse of a thin null
shell of matter, which may be represented by the Vaidya metric
\begin{equation}
ds^{2}=-\left(1-\frac{2M\theta(v)}{r}\right)dv^{2}+2\,dv\,dr+r^{2}d\Omega^{2}.\label{eq:vaidya}
\end{equation}
For $v>0$ the metric reduces to that of a Schwarzschild black hole
of mass $M$ and can be written in double-null form, 
\begin{equation}
ds^{2}=-\left(1-\frac{2M}{r}\right)du\,dv+r^{2}d\Omega^{2}.\label{eq:nullvaidya}
\end{equation}
The radial coordinate $r$ is expressed in terms of the null variables
$u$ and $v$ via
\begin{equation}
r(u,v)=2M\left(1+W\left(\exp\left(\frac{v-u}{4M}-1\right)\right)\right),\label{eq:lambert}
\end{equation}
where $W(x)$ is the Lambert function. 

We work in an approximation where the back-reaction on the black hole
geometry due to Hawking emission is ignored. Our goal is then to solve
the scalar equation of motion in the fixed background \eqref{eq:nullvaidya}
subject to smooth initial data, and then construct the quantum induced
stress tensor associated with this time-dependent solution. For $v>0$
the scalar equation of motion is simply:
\begin{equation}
\nabla^{2}\nabla^{2}\phi=\frac{(a+b)}{8\pi b}R_{\mu\nu\lambda\rho}R^{\mu\nu\lambda\rho}.\label{eq:scalareom-1}
\end{equation}
We will primarily be interested in the $T_{uu}$ component of the
stress tensor, which measures the outgoing flux on $\mathscr{I}^{+}$.
In the $(u,v)$ coordinate patch it evaluates to\begin{dmath}
\begin{equation}
T_{uu}=\frac{1}{6\pi r^{5}(2M-r)}\left(2r^{2}\left((2M-r)\phi^{(2,0)}(6aM+b(5M+r))+b\phi^{(1,1)}\left(8\pi r^{4}\phi^{(2,0)}+2M^{2}-5Mr+r^{2}\right)+br^{2}\left((r-8M)\phi^{(2,1)}+(2M-r)\phi^{(3,0)}-2r^{2}\phi^{(3,1)}\right)\right)+\phi^{(1,0)}\left(8\pi br^{4}\left(3(r-3M)\phi^{(1,1)}-5r^{2}\phi^{(2,1)}\right)+(2M-r)\left(4M^{2}(3a+4b)-4bMr+br^{2}\right)\right)+b\phi^{(0,1)}\left(4\pi r^{2}\left(\left(24M^{2}-16Mr+3r^{2}\right)\phi^{(1,0)}-6r^{2}(r-3M)\phi^{(2,0)}+2r^{4}\phi^{(3,0)}\right)-(2M-r)^{3}\right)-12\pi br^{2}(r-2M)^{2}\left(\phi^{(1,0)}\right)^{2}\right).\label{eq:Tuu}
\end{equation}
\end{dmath}Here we use the notation $\phi^{(n,m)}=\frac{\partial^{n}}{\partial u^{n}}\frac{\partial^{m}}{\partial v^{m}}\phi(u,v)$.

To approximate the condition there is vanishing ingoing energy flux
on $\mathscr{I}^{-}$, we impose the conditions $\phi=0$ and $\partial_{u}\phi=0$
at some large negative value of $u=u_{min}$. Likewise, to ensure
there is no quantum energy flux crossing the shock, we impose the
conditions $\phi=0$ and $\partial_{v}\phi=0$ at $v=0$. This leads
to a step function discontinuity in $\partial_{v}^{2}\phi$ at the
shock.

Restricting to a spherically symmetric scalar field $\phi(u,v)$,
the field equation \eqref{eq:scalareom-1} reduces to the following
partial differential equation in the patch $v>0$,
\begin{align}
 & 16\partial_{v}^{2}\partial_{u}^{2}\phi-\frac{16}{r}\left(1-\frac{3M}{r}\right)\left(\partial_{v}^{2}\partial_{u}-\partial_{u}^{2}\partial_{v}\right)\phi-\frac{16M^{2}}{r^{4}}\partial_{v}\partial_{u}\phi-\frac{4M}{r^{4}}\left(1-\frac{2M}{r}\right)^{2}\left(\partial_{v}-\partial_{u}\right)\phi\nonumber \\
= & \frac{6(a+b)M^{2}}{\pi br^{6}}\left(1-\frac{2M}{r}\right)^{2}\label{eq:eom}
\end{align}
where $r$ is given in terms of $u$ and $v$ by \eqref{eq:lambert}.

\subsection{Scaling relations}

To set up the numerics, it is helpful to define dimensionless coordinates
\begin{align}
u= & M\tilde{u}\,,\nonumber \\
v= & M\tilde{v}\,,\label{eq:rescaled}\\
r= & M\tilde{r}\,.\nonumber 
\end{align}
In terms of the tilded coordinates, $M$ scales out of \eqref{eq:eom}
and \eqref{eq:lambert}. Likewise, in these new coordinates, the stress
tensor rescales in a simple way. In particular,
\begin{equation}
T_{\tilde{u}\tilde{u}}=\frac{1}{M^{2}}\left(T_{uu}\left|_{M=1}\right.\right),\label{eq:rescaledTuu}
\end{equation}
so changing $M$ simply changes the normalization of the stress tensor.
Therefore, we can obtain the solution for all values of $M$ simply
by evaluating for $M=1$ and rescaling. Henceforth we will drop reference
to the tilded coordinates and simply scale to $M=1$.

It is also helpful to note that an overall rescaling of the dimensionless
constants $a$ and $b$ can be treated in a similar way. It leaves
the scalar equation of motion invariant, and linearly rescales the
stress tensor component in \eqref{eq:Tuu}. This leaves $a/b$ as
the only dimensionless parameter that can impact the evaluation of
the stress tensor beyond trivial rescaling. We will focus on the particular
case $a/b=1$ here but it is straightforward to scan through more
general values in a more comprehensive numerical treatment.

\subsection{Numerical Evaluation}

These equations are solved on a fixed grid in the $(u,v)$ plane using
the method of lines \citep{Lowe:1992ed,Lowe:1993ty}, in essence treating
each time-step in the $u$-direction as integrating a 2nd order ODE
in the $v$-direction with the shock boundary data determining the
solution. Likewise, each time-step in the $u-$direction can be viewed
as a 2nd order initial value problem with the $\mathscr{I}^{-}$ boundary
conditions. The code was implemented using Matlab. It was checked
that convergence was achieved by varying the grid size and changing
the desired accuracy tolerance of the code.

\begin{figure}
\includegraphics[scale=0.6]{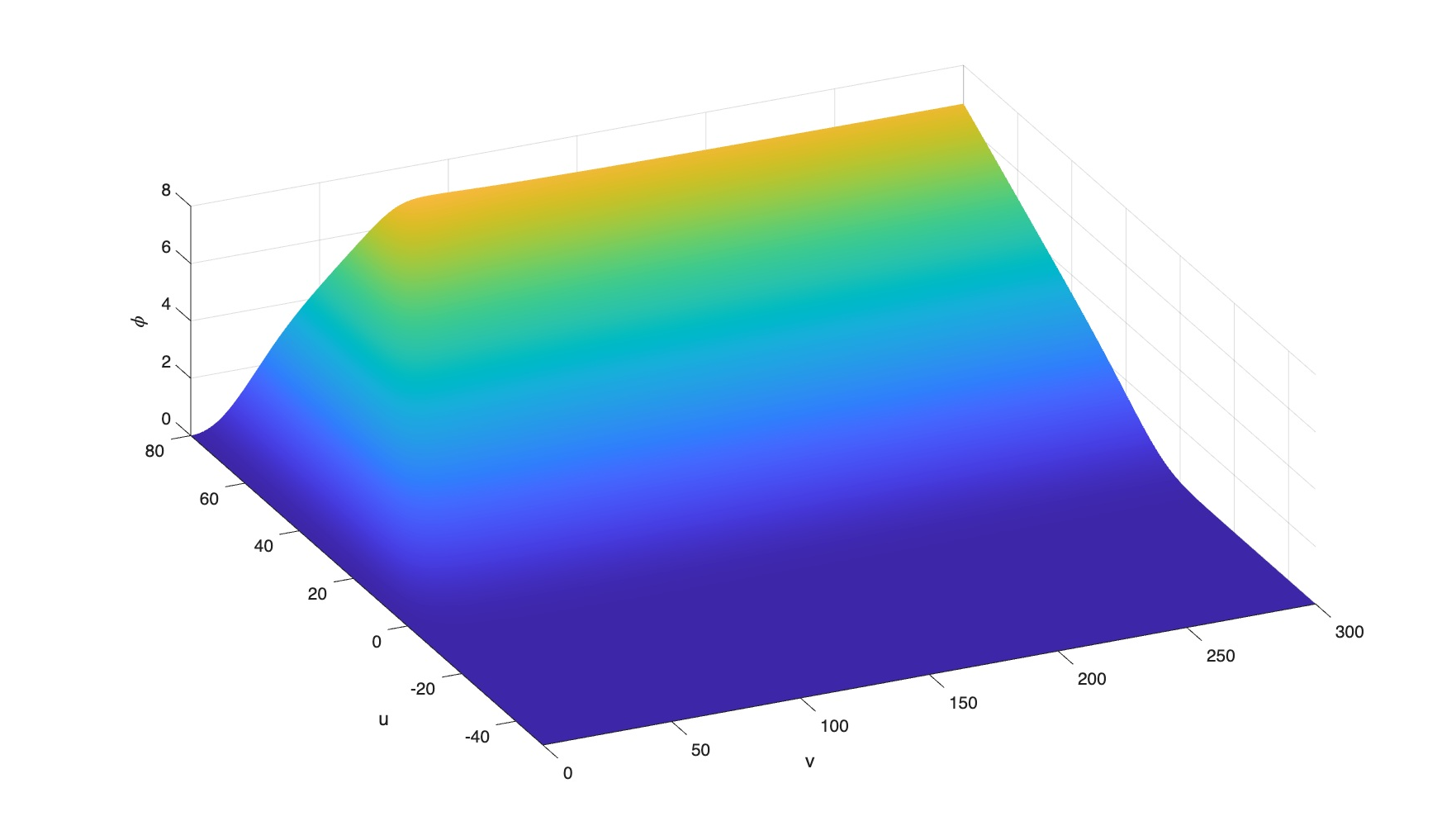}

\caption{The scalar field is plotted versus the null coordinates $u,\,v$.
The black hole is formed by a shock at $v=0$ the scalar field is
evolved out to $v_{max}=300$. The black hole horizon corresponds
to $u=+\infty$ at fixed $v$. The parameters are $M=1,\,a=1,\,b=1$.\label{fig:The-scalar-field}}

\end{figure}

The result for the scalar field profile is shown in figure \ref{fig:The-scalar-field},
where a simulation with $M=1,\,a=1,\,b=1$ is performed. Near $\mathscr{I}^{-}$
the field is fixed to vanish, but as one moves toward the horizon
for $u>0$ the curvature squared source in \eqref{eq:scalareom-1}
induces non-vanishing derivatives. Combined with the shock boundary
condition, this produces a time-dependent solution. Near $\mathscr{I}^{+}$
the solution appears to have a linear behavior in the $u$ coordinate
as we might expect from the Unruh solution found in \citep{Lowe:2025bxl}.
Likewise the solution appears to approach a constant near the horizon
as $v$ changes, indicating an approach to an outgoing wave.

\begin{figure}
\includegraphics[scale=0.6]{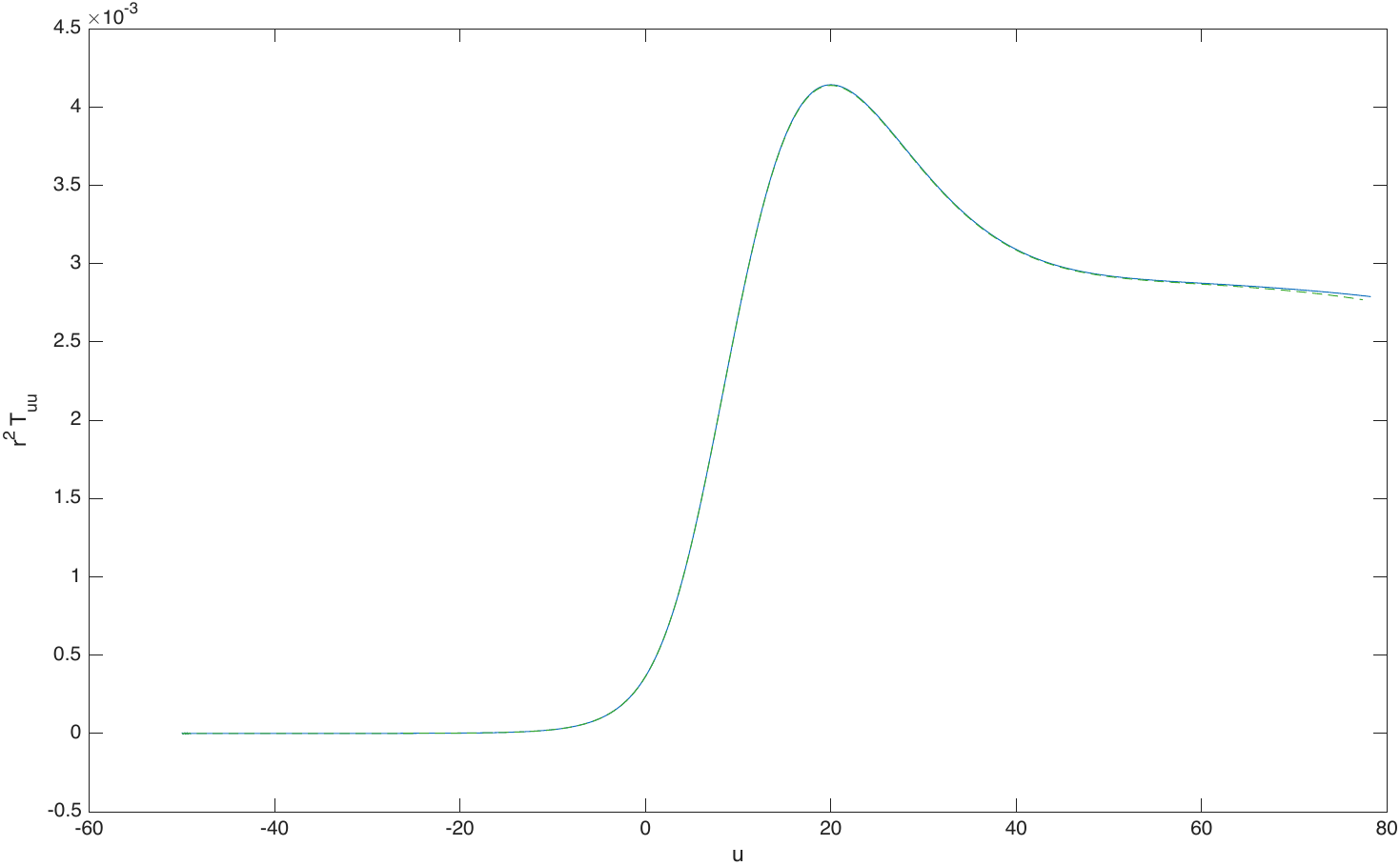}

\caption{The outgoing flux $T_{uu}$ is computed at $v=v_{max}$ to indicate
the flux of energy reaching $\mathscr{I}^{+}$. This does not turn
on until $u\approx0$ indicating emission from the region of horizon
formation. The parameters are $M=1,\,a=1,\,b=1$. The undashed line
shows the results for a fixed size grid with $(1.5\times10^{4})^{2}$
points, while the dashed line is for a grid with $(10^{4})^{2}$ points.
\label{fig:The-outgoing-flux}}

\end{figure}

The $T_{uu}$ component of the stress tensor, evaluated using \eqref{eq:Tuu},
multiplied by a factor of $r^{2}$ is shown in figure \ref{fig:The-outgoing-flux}.
This converges to this form for $v\gtrsim30$, indicating a $1/r^{2}$
falloff of $T_{uu}$ as expected for outgoing radiation. It is notable
that the outgoing energy flux vanishes as $u\to-\infty$, consistent
with the absence of the black hole before gravitational collapse.
There is then a positive pulse followed by a constant plateau. The
numerics becomes unstable for $u\apprge80$ due to transients formed
near the initial shock. Prior to the present calculation this feature
of Hawking radiation has only been precisely obtained in models of
two-dimensional gravity such as \citep{Callan:1992rs}. Heuristic
arguments that these 2d results can be extended to four dimensions
appear in \citep{Chen:2017pkl}. We comment further on the late time
behavior in the following section.

\section{Semi-Analytic Late-Time Solution}

Near $\mathscr{I}^{+}$ (i.e. $v\to\infty,$ fixed $u$), the scalar
field appears to approach a solution of the form
\begin{equation}
\phi=\alpha u+\beta\,,\label{eq:latetimephi}
\end{equation}
provided $u\gtrsim0$, as can be seen in figure \ref{fig:The-scalar-field}.
Using the results of \citep{Lowe:2025bxl}, we can evaluate the quantum
stress energy tensor component corresponding to outgoing radiation
for this solution
\begin{equation}
T_{uu}=\frac{2b\alpha^{2}}{r^{2}}\,,\label{eq:latetimeTuu}
\end{equation}
and see that $r^{2}T_{uu}$ does indeed approach a positive constant.
In this semi-analytic solution, the slope parameter $\alpha$ must
be determined from the dynamical solution at $u\gtrsim0$. The scalar
field does not approach linear behavior for $u<0$, before the onset
of Hawking emission.

The Unruh-like solution found in \citep{Lowe:2025bxl} has the form
\begin{equation}
\phi=\frac{(a+b)}{8\pi Mb}u+\cdots,\label{eq:unruhlike}
\end{equation}
yielding an outgoing flux
\begin{equation}
T_{uu}=\frac{(a+b)^{2}}{32\pi^{2}M^{2}br^{2}},\label{eq:unruhflux}
\end{equation}
From the simulation, for $M=1$, $a=1$, $b=1$ we see that $\alpha=0.083$
while analytically we find $\alpha=0.080$ for the static Unruh solution.
We take this as evidence the system is approaching a quasi-static
solution given by the solution of \citep{Lowe:2025bxl}.

\section{Discussion}

The present work provides the first time-dependent treatment of Hawking
radiation in four spacetime dimensions. By numerically evolving an
auxiliary scalar across a genuine collapse spacetime, rather than
a stationary Schwarzschild geometry, we have demonstrated that the
semiclassical dynamics does not produce outgoing energy flux until
the matter has undergone the gravitational collapse. This resolves
a long-standing conceptual issue, namely the appearance of \textquotedblleft pre-Hawking\textquotedblright{}
flux in static states. Our results establish that this artifact is
not intrinsic to the effective action itself, but arises solely from
imposing a static state across a geometry that should instead be sourced
dynamically.

The switch-on behavior we obtain is qualitatively similar to that
seen analytically in the CGHS model \citep{Callan:1992rs}, even though
in four dimensions there is no exact analytic reduction. The agreement
strongly suggests that the essential mechanism behind the onset of
Hawking radiation---namely, the localization of vacuum polarization
effects to the near-horizon region and the subsequent leaking of energy
through the outgoing null channel---is universal across semiclassical
theories. In particular, the fact that the effective action \eqref{eq:action}
reproduces not only the trace anomaly, but also the correct dynamical
ignition of outgoing flux, indicates that the auxiliary scalar does
indeed encode physical degrees of freedom responsible for horizon-scale
vacuum polarization.

An essential implication of our result is that the anomaly-induced
effective theory remains predictive when treated as an initial-value
problem. In the static analysis of \citep{Lowe:2025bxl}, regularity
at the future horizon and vanishing ingoing energy flux fixed a unique
Unruh-like state. Here, the same late-time configuration is reached
through smooth initial conditions on $\mathcal{\mathscr{I}}^{-}$
without the need to impose horizon boundary data by hand. This unique
state therefore appears as an attractor under time evolution. In this
sense, the effective action is robust to choice of initial data: any
smooth infalling configuration forming a horizon should asymptote
to the same semiclassical radiating solution.

Future work must address the semiclassical back-reaction problem.
Since the outgoing flux is now shown to arise dynamically from the
collapse geometry itself, the next step is to close the system by
evolving the metric under the semiclassical Einstein equation with
anomaly-induced stress tensor on the right-hand side. It is precisely
at this point that the non-linear instability hinted at by the static
analysis \citep{Lowe:2025bxl} becomes relevant. Presumably Hawking
evaporation will steadily reduce the mass parameter, until at late
time the semiclassical approximation fails. The theory may either
settle gradually into a remnant configuration, encounter a sharp strongly
coupled Planckian endpoint, or transition to some qualitatively different
quantum gravitational phase.

More broadly, the present framework opens the door to resolving two
related questions: 
\begin{itemize}
\item Late-time breakdown and endpoint physics: Does the anomaly-induced
effective theory predict a finite evaporation lifetime, and if so,
does black hole evaporation terminate smoothly or via singular breakdown?
\item No-hair consistency in dynamical settings: Does dynamical evolution
always eliminate auxiliary scalar hair, as in the static case \citep{Lowe:2025bxl},
or can transient configurations imprint observable effects on intermediate-time
evaporation?
\end{itemize}
By demonstrating that an anomaly-induced auxiliary scalar field generates
Hawking flux only subsequent to gravitational collapse, we have addressed
a major conceptual deficiency in semiclassical constructions and provided
the first account of the onset of Hawking emission in a four-dimensional
effective field theory. The qualitative agreement with the CGHS emission
profile indicates that the onset of Hawking emission is not an accident
of two-dimensional conformal structure but a universal feature of
conformal anomaly induced semiclassical dynamics. Incorporating semiclassical
back-reaction constitutes a natural continuation of this work. Our
dynamical framework provides a controlled setting for the study black
hole evaporation in four spacetime dimensions. 
\begin{acknowledgments}
The work of L.T. is supported in part by the Icelandic Research Fund
grant 228952-053. 
\end{acknowledgments}

\bibliographystyle{utcaps}
\bibliography{riegert}

\providecommand{\href}[2]{#2}\begingroup\raggedright\begin{thebibliography}{10}

\bibitem{Hawking1975}
S.~W. Hawking, ``Particle creation by black holes,''
  \href{https://dx.doi.org/10.1007/BF02345020}{{\em Communications in
  Mathematical Physics} {\bfseries 43} no.~3, (Aug, 1975) 199--220}.
  \url{https://link.springer.com/content/pdf/10.1007/BF02345020.pdf}.

\bibitem{Callan:1992rs}
C.~G. Callan, Jr., S.~B. Giddings, J.~A. Harvey, and A.~Strominger,
  ``{Evanescent black holes},''
  \href{https://dx.doi.org/10.1103/PhysRevD.45.R1005}{{\em Phys. Rev. D}
  {\bfseries 45} no.~4, (1992) R1005},
  \href{https://arxiv.org/abs/hep-th/9111056}{{\ttfamily
  arXiv:hep-th/9111056}}.

\bibitem{Russo:1992ax}
J.~G. Russo, L.~Susskind, and L.~Thorlacius, ``{The Endpoint of Hawking
  radiation},'' \href{https://dx.doi.org/10.1103/PhysRevD.46.3444}{{\em Phys.
  Rev. D} {\bfseries 46} (1992) 3444--3449},
  \href{https://arxiv.org/abs/hep-th/9206070}{{\ttfamily
  arXiv:hep-th/9206070}}.

\bibitem{RIEGERT198456}
R.~J. Riegert, ``A non-local action for the trace anomaly,''
  \href{https://dx.doi.org/https://doi.org/10.1016/0370-2693(84)90983-3}{{\em
  Physics Letters B} {\bfseries 134} no.~1, (1984) 56--60}.
  \url{https://www.sciencedirect.com/science/article/pii/0370269384909833}.

\bibitem{Balbinot:1999ri}
R.~Balbinot, A.~Fabbri, and I.~L. Shapiro, ``{Anomaly induced effective actions
  and Hawking radiation},''
  \href{https://dx.doi.org/10.1103/PhysRevLett.83.1494}{{\em Phys. Rev. Lett.}
  {\bfseries 83} (1999) 1494--1497},
  \href{https://arxiv.org/abs/hep-th/9904074}{{\ttfamily
  arXiv:hep-th/9904074}}.

\bibitem{Mottola:2006ew}
E.~Mottola and R.~Vaulin, ``{Macroscopic Effects of the Quantum Trace
  Anomaly},'' \href{https://dx.doi.org/10.1103/PhysRevD.74.064004}{{\em Phys.
  Rev. D} {\bfseries 74} (2006) 064004},
  \href{https://arxiv.org/abs/gr-qc/0604051}{{\ttfamily arXiv:gr-qc/0604051}}.

\bibitem{Mottola:2016aa}
E.~Mottola, ``{Scalar Gravitational Waves in the Effective Theory of
  Gravity},'' \href{https://dx.doi.org/10.1007/JHEP07(2017)043}{{\em JHEP}
  {\bfseries 07} (2017) 043},
  \href{https://arxiv.org/abs/1606.09220}{{\ttfamily arXiv:1606.09220
  [gr-qc]}}. [Erratum: JHEP 09, 107 (2017)].

\bibitem{Mottola:2025fhl}
E.~Mottola, ``{Gravitational vacuum condensate stars in the effective theory of
  gravity},'' \href{https://dx.doi.org/10.1103/PhysRevD.111.104018}{{\em Phys.
  Rev. D} {\bfseries 111} no.~10, (2025) 104018},
  \href{https://arxiv.org/abs/2502.02519}{{\ttfamily arXiv:2502.02519
  [gr-qc]}}.

\bibitem{Lowe:2025bxl}
D.~A. Lowe and L.~Thorlacius, ``{Effective field theory description of Hawking
  radiation},'' \href{https://dx.doi.org/10.1007/JHEP11(2025)057}{{\em JHEP}
  {\bfseries 11} (2025) 057},
  \href{https://arxiv.org/abs/2505.07722}{{\ttfamily arXiv:2505.07722
  [hep-th]}}.

\bibitem{Christensen:1977jc}
S.~M. Christensen and S.~A. Fulling, ``{Trace Anomalies and the Hawking
  Effect},'' \href{https://dx.doi.org/10.1103/PhysRevD.15.2088}{{\em Phys. Rev.
  D} {\bfseries 15} (1977) 2088--2104}.

\bibitem{POLYAKOV1981207}
A.~Polyakov, ``Quantum geometry of bosonic strings,''
  \href{https://dx.doi.org/https://doi.org/10.1016/0370-2693(81)90743-7}{{\em
  Physics Letters B} {\bfseries 103} no.~3, (1981) 207--210}.
  \url{https://www.sciencedirect.com/science/article/pii/0370269381907437}.

\bibitem{Duff:1977ay}
M.~J. Duff, ``{Observations on Conformal Anomalies},''
  \href{https://dx.doi.org/10.1016/0550-3213(77)90410-2}{{\em Nucl. Phys. B}
  {\bfseries 125} (1977) 334--348}.

\bibitem{Duff:1993wm}
M.~J. Duff, ``{Twenty years of the Weyl anomaly},''
  \href{https://dx.doi.org/10.1088/0264-9381/11/6/004}{{\em Class. Quant.
  Grav.} {\bfseries 11} (1994) 1387--1404},
  \href{https://arxiv.org/abs/hep-th/9308075}{{\ttfamily
  arXiv:hep-th/9308075}}.

\bibitem{Liu:2025xfu}
B.-N. Liu, D.~A. Lowe, and L.~Thorlacius, ``{Generalized Effective Field Theory
  for Four-Dimensional Black Hole Evaporation},''
  \href{https://arxiv.org/abs/2511.05374}{{\ttfamily arXiv:2511.05374
  [hep-th]}}.

\bibitem{Lowe:1992ed}
D.~A. Lowe, ``{Semiclassical approach to black hole evaporation},''
  \href{https://dx.doi.org/10.1103/PhysRevD.47.2446}{{\em Phys. Rev. D}
  {\bfseries 47} (1993) 2446--2453},
  \href{https://arxiv.org/abs/hep-th/9209008}{{\ttfamily
  arXiv:hep-th/9209008}}.

\bibitem{Lowe:1993ty}
D.~A. Lowe, {\em {Aspects of two-dimensional quantum gravity}}.
\newblock Umi-94-07117, 11, 1993.

\bibitem{Chen:2017pkl}
P.~Chen, W.~G. Unruh, C.-H. Wu, and D.-H. Yeom, ``{Pre-Hawking radiation cannot
  prevent the formation of apparent horizon},''
  \href{https://dx.doi.org/10.1103/PhysRevD.97.064045}{{\em Phys. Rev. D}
  {\bfseries 97} no.~6, (2018) 064045},
  \href{https://arxiv.org/abs/1710.01533}{{\ttfamily arXiv:1710.01533
  [gr-qc]}}.

\end{thebibliography}\endgroup

\end{document}